\documentclass{elsart}
\usepackage{graphicx}

\begin{document}
\begin{frontmatter}
\title{Fine structure of Vavilov-Cherenkov radiation
near the Cherenkov threshold}
\author[jinr,IThF]{V.G. Kartavenko\thanksref{corr}},
\author[jinr]{G.N. Afanasiev\thanksref{corr}} and
\author[IThF]{W. Greiner}
\address[jinr]{Bogoliubov Laboratory of Theoretical Physics,
Joint Institute for Nuclear Research, Dubna, Moscow District, 141980,
Russia}
\address[IThF]{Institut f\"ur Theoretische Physik der J. W. Goethe
Universit\"at\\
D-60054 Frankfurt am Main, Germany}
\thanks[corr]{Corresponding author. Phone: +7 09621 63710;
Fax: +7 09621 65084;\\
\hspace*{43mm}E-mail: kart@thsun1.jinr.ru}

\begin{abstract}
We analyze the Vavilov-Cherenkov radiation (VCR)
in a dispersive nontransparent dielectric air-like
medium both below and above the Cherenkov threshold,
in the framework of classical electrodynamics.
It is shown that the transition to the subthreshold energies
leads to the destruction of electromagnetic shock waves and
to the sharp reduction of the frequency domain where VCR is emitted.
The fine wake-like structure of the Vavilov-Cherenkov radiation
survives and manifests the existence of the
subthreshold radiation in the domain of anomalous dispersion.
These domains can approximately be defined by the two
phenomenological parameters of the medium,
namely, the effective frequency of oscillators and the damping
describing an interaction with the other degrees of freedom.\\[3mm]
\end{abstract}

\end{frontmatter}
\newpage
\section{Introduction}
Vavilov-Cherenkov radiation (VCR) is a well established phenomenon
widely used in physics and technology~\cite{Frank88,Greiner98}.
Nevertheless,  some of its fundamental aspects are still open.
Recently, interest in the problem of surpassing
the Cherenkov threshold has
been renewed. A group of researchers from Comenius Univercity,
Bratislava, CERN and JINR have carried out measurements of VCR
induced by a high-energy beam of lead ions in air, helium
and various crystals (NA49 experiment)~\cite{CC99}.
The members of this group supposed~\cite{ZrelovTR98},
using the previously obtained experimental data on VCR induced
by relativistic electrons in air~\cite{Ruzicka},
that the pre-Cherenkov
electromagnetic shock wave could arise when the charge
particle velocity $v$ coincides with
the phase velocity of light in the medium
$c_n=c/n_0$  ($n_0$ is the refractive index of the medium,
and $c$ denotes light velocity in the vacuum).
For a relativistic electron moving in a gas
the radiation intensity was measured as a function of
a gas pressure $P$, which is related in a simple way to
the gas refractive index $n_0$.
Changing continuously the refractive index of the medium one gets
a tool to analyse VCR both below and above the Cherenkov threshold.
A small but finite value of intensity
indicates the existence of VCR
below the threshold velocity $v<c_n$~\cite{Ruzicka}.
In spite of the fact that these experiments are preliminary,
it is useful to conduct a detailed theoretical analysis
of VCR both below and above the Cherenkov threshold.
It is the goal of this paper.

We investigate electromagnetic fields (EMF)
radiated by a point
charge uniformly moving  in a dielectric dispersive medium.
It is suggested that uniform motion of a particle is maintained by
some external force whose origin is not of interest for us.
In the case of frequency independent electric permittivity
$\epsilon_0=\sqrt{n_0}$,
there is no electromagnetic field (EMF)
before the Cherenkov cone accompanying the charge,
an infinite EMF on the Cherenkov cone itself and finite values
of EMF strengths behind
it~\cite{Heaviside12,TammFrank37}.
So information concerning
the transition effects arising when  charge velocity
coincides with $c_n$ is lost (except for the existence of
the Cherenkov shock wave itself)~\cite{AfanasievES98}.

In the classical electrodynamics~\cite{Greiner98,AhSh93}
VCR is a collective phenomenon involving a large number of
atoms of the medium whose electrons are excited by EMF
of the passing particle. Therefore, we use
the macroscopic complex dielectric function $\epsilon$
~\cite{AhSh93,Fermi40}
\begin{equation}
\epsilon(\omega,p)=1+\frac{\omega_L^2}{\omega_0^2-\omega^2+ip\omega}.
\label{eps}
\end{equation}
This is the standard parameterization
describing many optical phenomena~\cite{BornWolf75} and
is well suited for air-like media
investigated in~\cite{ZrelovTR98}.
Following~\cite{Brillouin60,LagTig96},
we extrapolate the parameterization
(\ref{eps}) to all $\omega$. This means that we disregard the
excitation of nuclear levels and the discrete structure of scatterers.

\section{Electromagnetic fields}
\begin{figure}[h,t,b]
\vspace*{-14mm}
\begin{center}
\hspace*{-13mm}
\includegraphics[angle=-90,width=84mm]{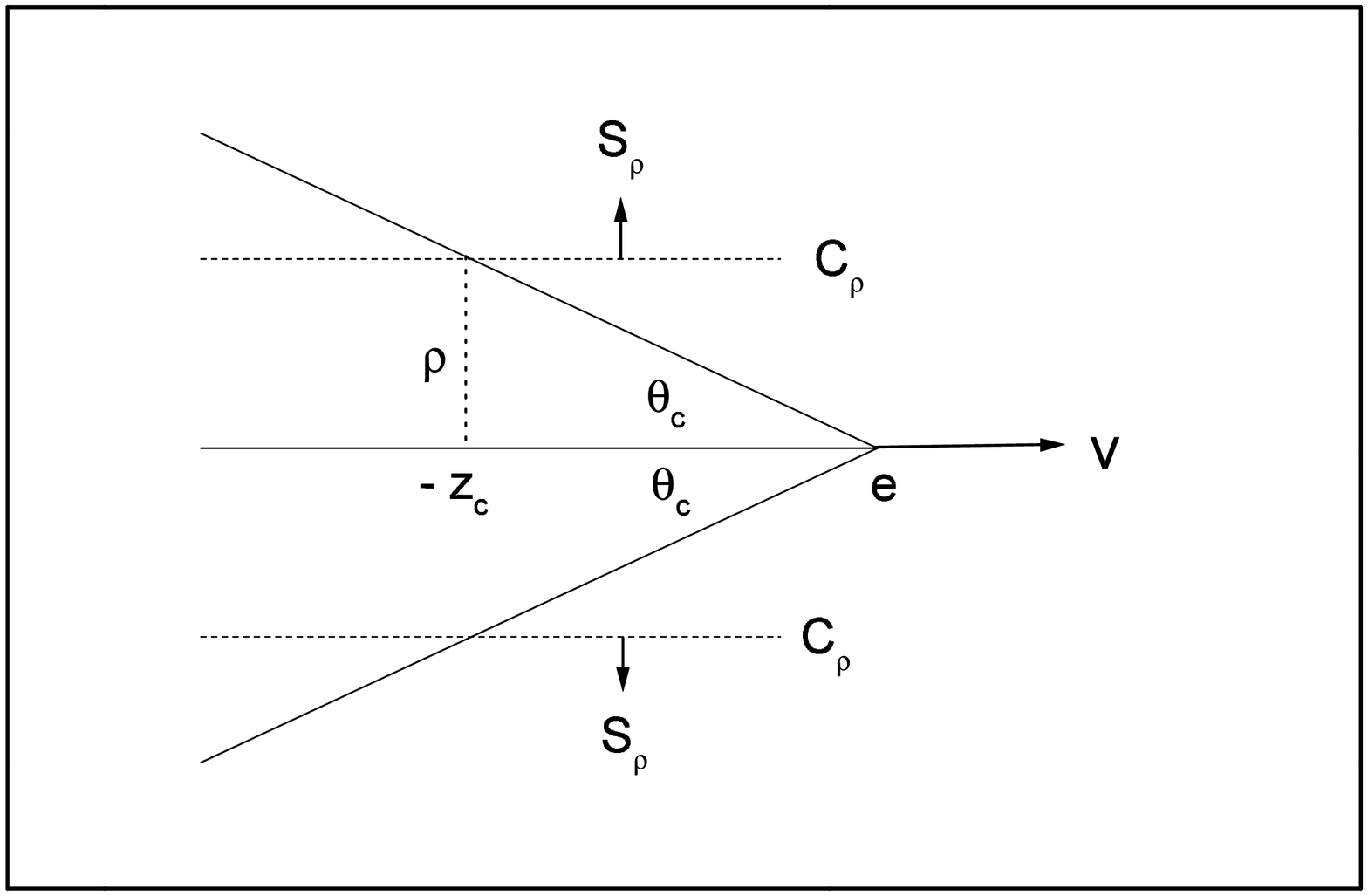}      
\end{center}
\vspace*{-5mm}
\caption{\small The problem under consideration: an electron moves
in a medium in positive $z$-direction with constant velocity $v$
($\rho$ and $z$ are cylindrical coordinates in the projectile rest
frame). The Cherenkov cone is presented schematically
for a medium of a constant dielectric coeficient.
$\theta_c=\arcsin(c_n/v)$ is the Cherenkov angle.
 On the surface of the cylinder C$_\rho$
EMF is zero for $z>-z_c$. The component $S_\rho$ of Poynting's vector
describes  EMF energy flux through the surface of the cylinder,
which confined  to the surface of the cone.
EMF inside the cone does not contribute
to the radiation.}
\label{fig:problem}
\end{figure}

The charge and  current densities
of the treated
problem, Fig.~\ref{fig:problem}, are given by\footnote{We follow
the original papers of Tamm-Frank~\cite{TammFrank37} and
Fermi~\cite{Fermi40} to treat VCR. The
important peculiarities of the VCR near the Cherenkov threshold
due to a finite thickness of
the radiator~\cite{Tamm39,Pafomov57,Aitken63,KobzevFrank81} and
possible effects due to accelerated and
decelerated motion~\cite{AfanasievES98}
will be considered separately.}
$\rho(\vec r,t)=e\delta(x)\delta(y)\delta(z-vt),\; j_z=v\rho.$
The cylindrical components of EMF strengths
can be written in the following form~\cite{AKjpd98}
\begin{eqnarray} \nonumber
H_\phi &=&\frac{2e}{\pi v c}\int\limits_{0}^{\infty}
d\omega\omega|1-\beta^2\epsilon|^{1/2}\Bigl(
\bigl(\cos(\frac{\phi}{2}+\alpha) K_{1r}
-\sin( \frac{\phi}{2}+\alpha) K_{1i}\Bigr),\\ \nonumber
E_z &=&-\frac{2}{\pi v^2}\int\limits_{0}^{\infty}
d\omega\omega
\biggl(\Bigl( \cos\alpha(\epsilon_r^{-1}-\beta^2)-
\epsilon_i^{-1}\sin\alpha \Bigr) K_{0i}\\ \nonumber
 &+& \Bigl(\sin\alpha(\epsilon_r^{-1}-\beta^2)+
\epsilon_i^{-1}\cos\alpha \Bigr) K_{0r}\biggr),\\ \nonumber
E_\rho &=&\frac{2}{\pi v^2}\int\limits_{0}^{\infty}
d\omega\omega|1-\beta^2\epsilon|^{1/2}
\Bigl( (\epsilon_r^{-1} \cos\alpha-
\epsilon_i^{-1}\sin\alpha)
(\cos\frac{\phi}{2}K_{1r}-\sin\frac{\phi}{2}K_{1i})\\
 &-& (\epsilon_i^{-1}\cos\alpha+\epsilon_r^{-1}\sin\alpha)
(\sin\frac{\phi}{2}K_{1r}+\cos\frac{\phi}{2}K_{1i})\Bigr).
\label{cfields}
\end{eqnarray}
\begin{eqnarray}\nonumber
K_{mr}&=&Re K_m(\frac{\rho\omega}{v}\sqrt{1-\beta^2\epsilon}),
\quad
K_{mi}=Im K_m(\frac{\rho\omega}{v}\sqrt{1-\beta^2\epsilon}),
\quad{\mathrm m=0,1}\\  \nonumber
\alpha&\equiv&\omega(t-z/v),\quad
\epsilon_r\equiv Re \epsilon,\quad
\epsilon_i\equiv Im \epsilon,\\ \nonumber
\phi&\equiv&\arg\left(1-\beta^2\epsilon\right),\quad
Re\sqrt{1-\beta^2\epsilon} > 0.
\end{eqnarray}
We present here only a minimal set of mathematical relations
to evaluate EMF. For mathematical details (derivation of
equations, analysis of possible divergences,
semiclassical WKB estimates, gauge and polarization selections,
Kronig-Kramers dispersion relations etc.) we refer
to~\cite{AKjpd98,AKM98}.

\begin{figure}[h,b]
\vspace*{1mm}
\begin{center}
\hspace*{-15mm}
\includegraphics[width=70mm]{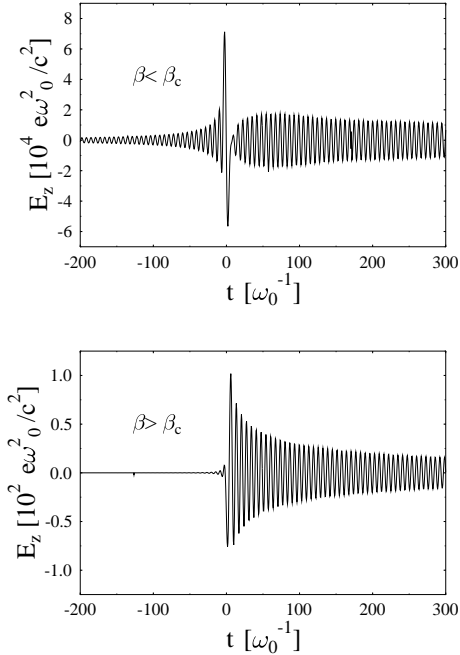}     
\end{center}
\vspace*{-5mm}
\caption{\small Time dependence of the
electric field strength on
the surface of the cylinder C$_\rho$ both below and
above Cherenkov threshold.}
\label{fig:ez}
\vspace*{-38mm}
\end{figure}

\begin{figure}[h]
\vspace*{-1mm}
\begin{center}
\hspace*{-10mm}
\includegraphics[width=70mm]{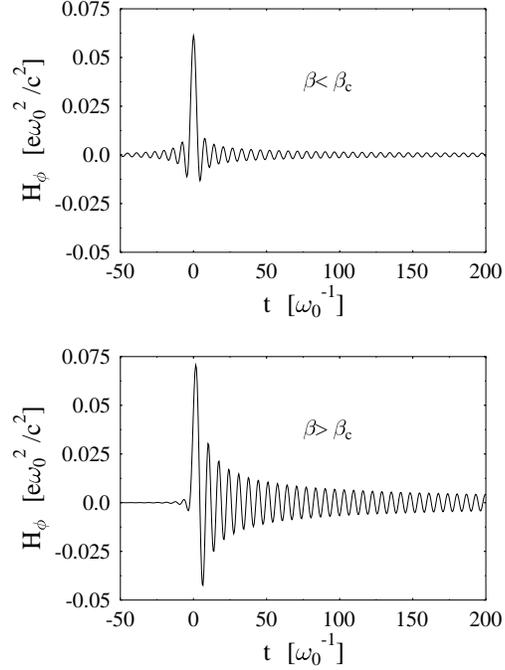}           
\end{center}
\vspace*{-5mm}
\caption{\small Time dependence of the magnetic field strength
on the surface of the cylinder C$_\rho$.}
\label{fig:hfi}
\end{figure}

The time dependence of EMF strengths at the surface of
the cylinder C$_\rho$ ($\rho=10\,[c/\omega_0],\;z=0$),
induced by a fast particle moving along the "z" axis
is exhibited in Figs.~\ref{fig:ez} and \ref{fig:hfi}.
To describe  the passing of a light barrier by a charge particle
in the experiments~\cite{ZrelovTR98},
we select the following parameters.
The pressure is related to the gas density $N_g$
by the well-known thermodynamic relation: $PV=k N_g T$,
where $V$ is the gas volume, $T$ is its temperature and $k $ is
the Boltzmann constant.
In the definition of $\epsilon$ (\ref{eps})
$\omega_L$ is the plasma
frequency $\omega_L^2=4\pi N_e e^2/m$,
$N_e=N_gZ$ is the number of electrons
per unit of volume, $m$ is the electron mass,
$\omega_0$ is some resonance frequency and
$Z$ is the atomic number of the gas particles.
The particle velocity is $\beta\sim~0.999$.
The Cherenkov threshold can be associated with the critical
velocity $\beta_c\equiv\sqrt{1/\epsilon(0,0)}$~\cite{AKjpd98}.
Changing $\beta_c$  from
$0.99$ to $0.9999$, describes a transition from
the superlight regime to the sublight one,
and corresponds to an approximately
100 time decrease in the gas pressure.

In Figs.~\ref{fig:ez} and \ref{fig:hfi}
one can see formation of the well-defined front of electromagnetic
shock waves for $\beta>\beta_c$ and suppression of a pre-front
wave structure  which existed for $\beta<\beta_c$.

A fine structure of EMF strengths presented in
Figs.~\ref{fig:ez} and \ref{fig:hfi}
indicates that the response of
a dispersive medium to a penetrating charge particle
manifests itself in an axially symmetrical wake-like~\cite{wake}
space-time distribution. These are well-known properties,
in the case of $\beta>\beta_c$, of any
homogeneous isotropic electron plasma with a dielectric
function $\epsilon(\omega,0)
$~\cite{BohmPines,shock-solids2,shock-solids1}.
We will use the terminology of ``wake'', although the Cherenkov
cone does not exist for subthreshold velocities  $\beta<\beta_c$.

\section{The radiated EMF energy flux}
\begin{figure}[h,t,b]
\vspace*{-1mm}
\begin{center}
\hspace*{-10mm}
\includegraphics[width=70mm]{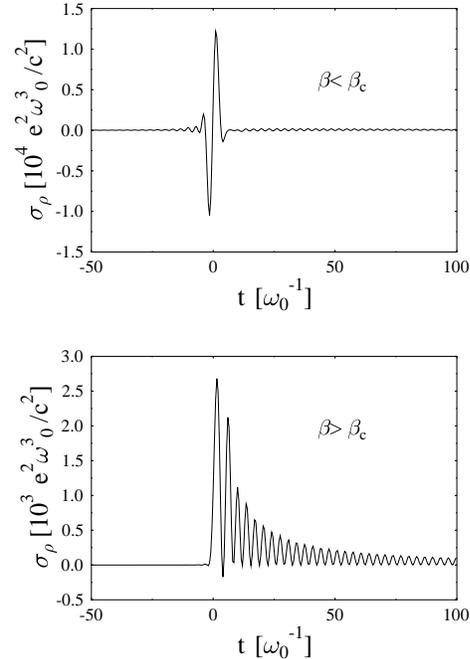}            
\end{center}
\vspace*{-5mm}
\caption{\small Time dependence of the radiated EMF energy flux
per unit length of the surface of the cylinder C$_\rho$
both below and above Cherenkov threshold.}
\label{fig:sro}
\end{figure}

The energy flux per unit  length radiated through the surface
of a cylinder C$_\rho$ (Fig.~\ref{fig:problem})
coaxial with the $z$ axis for the total time of motion is given by
Poynting's vector $S_\rho=-(c/4\pi)E_z H_\phi$
\begin{eqnarray}
W_{\rho}=\int\limits_{-\infty}^{+\infty}\sigma_\rho dt=
\frac{1}{v}\int\limits_{-\infty}^{+\infty}\sigma_\rho dz ,\quad
 \sigma_\rho=2\pi\rho S_\rho.
\label{W}
\end{eqnarray}
Time dependence of EMF flux densities $\vec{\sigma}(t)$
is presented in
Fig.~\ref{fig:sro}.
One can see very different time distributions
for subthreshold and normal regimes.
The EMF flux is localized more strongly than the EMF strengths.
It is a result of complicated interference of  electric
and magnetic strengths entering via Eq~(\ref{W}) in the
definition of Poynting's vector.

Using $E_z$ and $H_\phi$ given by (\ref{cfields})
for the whole time of charge motion
one gets
$W_{\rho}=\int\limits_{0}^{\infty}f_{\rho}(\omega) d\omega,$
where the $\rho$ component of a spectral density
is given by \cite{AKjpd98}
\begin{eqnarray}\nonumber
f_{\rho}(\omega)=&-&\frac{2e^2\rho}{\pi v^3}\omega^2
|1-\beta^2\epsilon|^{1/2}\times \\ \nonumber
&\times& \biggl( (K_{0r}K_{1r}+K_{0i}K_{1i})
\Bigl((\epsilon_r^{-1}-\beta^2)\sin
\frac{\phi}{2}-\epsilon_i^{-1}\cos\frac{\phi}{2} \Bigr)\\
&-&(K_{0i}K_{1r}-K_{0r}K_{1i})
\Bigl( (\epsilon_r^{-1}-\beta^2)
\cos\frac{\phi}{2}+\epsilon_i^{-1}\sin\frac{\phi}{2}\Bigr)\biggr).
\label{fw}
\end{eqnarray}
At a large distance in the limit $p\to 0$ this expression turns into
the well-known relations~\cite{TammFrank37}
\begin{eqnarray}\nonumber
\lim_{p\to 0}f_{\rho}(\omega) &=& 0,\quad{\mathrm for}\;
\beta^2\epsilon<1,\\ \nonumber
\lim_{p\to 0}f_{\rho}(\omega) &=&
\frac{e^2\omega}{c^2}(1-\frac{1}{\epsilon\beta^2}),
\quad{\mathrm for}\,\beta^2\epsilon>1,\\
\lim_{p\to 0} W_{\rho} &=&
\frac{e^2}{c^2}\int\limits_{\beta^2\epsilon>1}^{}\omega
d\omega (1-\frac{1}{\epsilon\beta^2}).
\label{w0}
\end{eqnarray}
The validity of Eq.~(\ref{w0}) has been confirmed
in~\cite{Fermi40,Ginzburg96}.

\begin{figure}[h,t,b]
\begin{center}
\hspace*{-12mm}
\includegraphics[width=70mm]{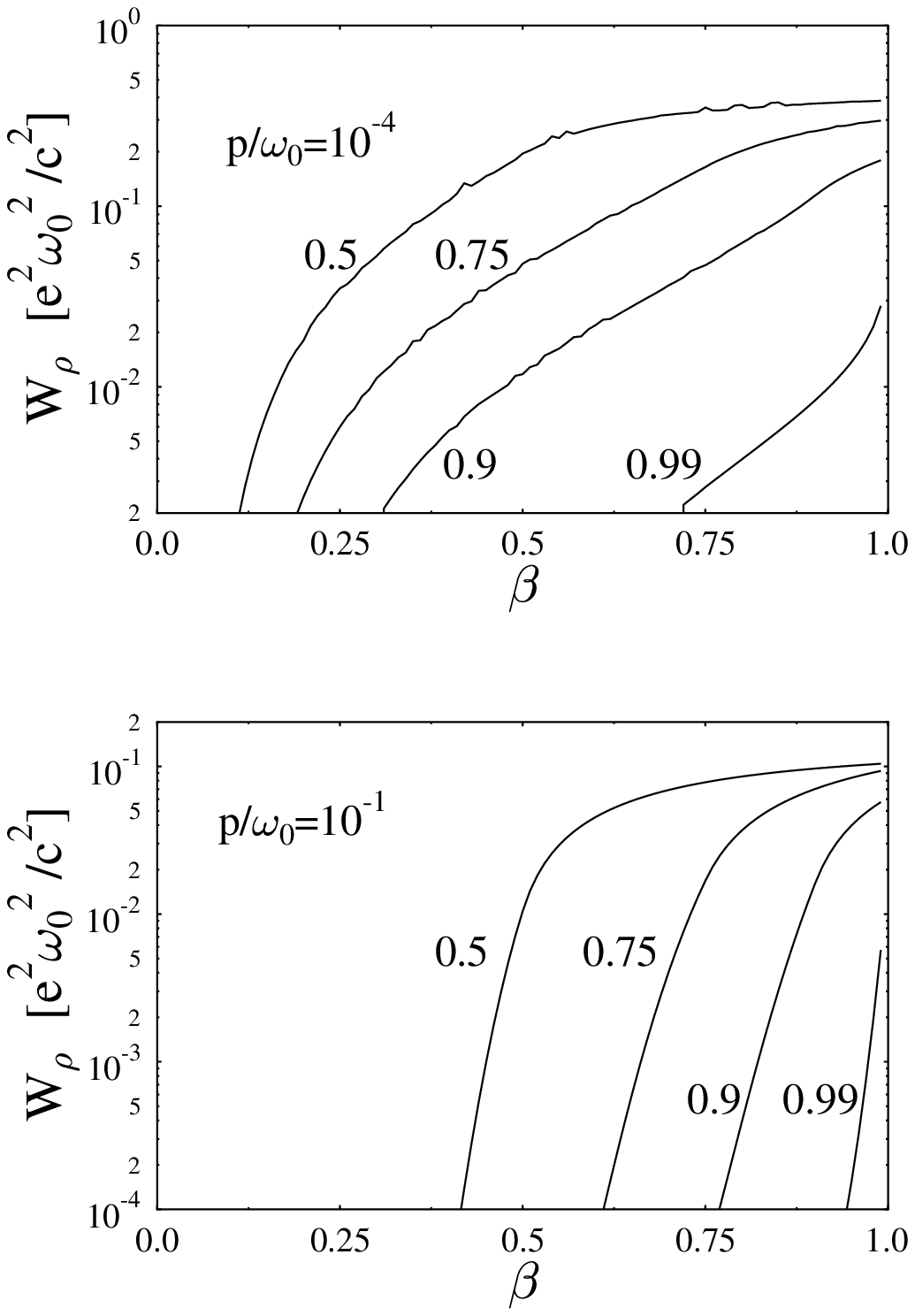}         
\end{center}
\vspace*{-5mm}
\caption{\small The radiated energy losses per unit length as
a function of a electron velocity for two damping parameters.
The numbers on a particular line means $\beta_c$ (a static
refractive index of a media $n_0=1/\beta_c$).}
\label{fig:fb}
\end{figure}

In Fig.~\ref{fig:fb},
we present the radiated energy losses
per unit length $W_\rho$ on a cylinder C$_\rho$  as a function
of a particle velocity for a few parameters of $\beta_c$. One can
see that for the fixed permittivity of the medium there exist
radiated energy losses for the subthreshold velocities.
The increasing damping reduces domains of
a particle velocity with an essential VCR.

\begin{figure}[h]
\begin{center}
\hspace*{-12mm}
\includegraphics[width=70mm]{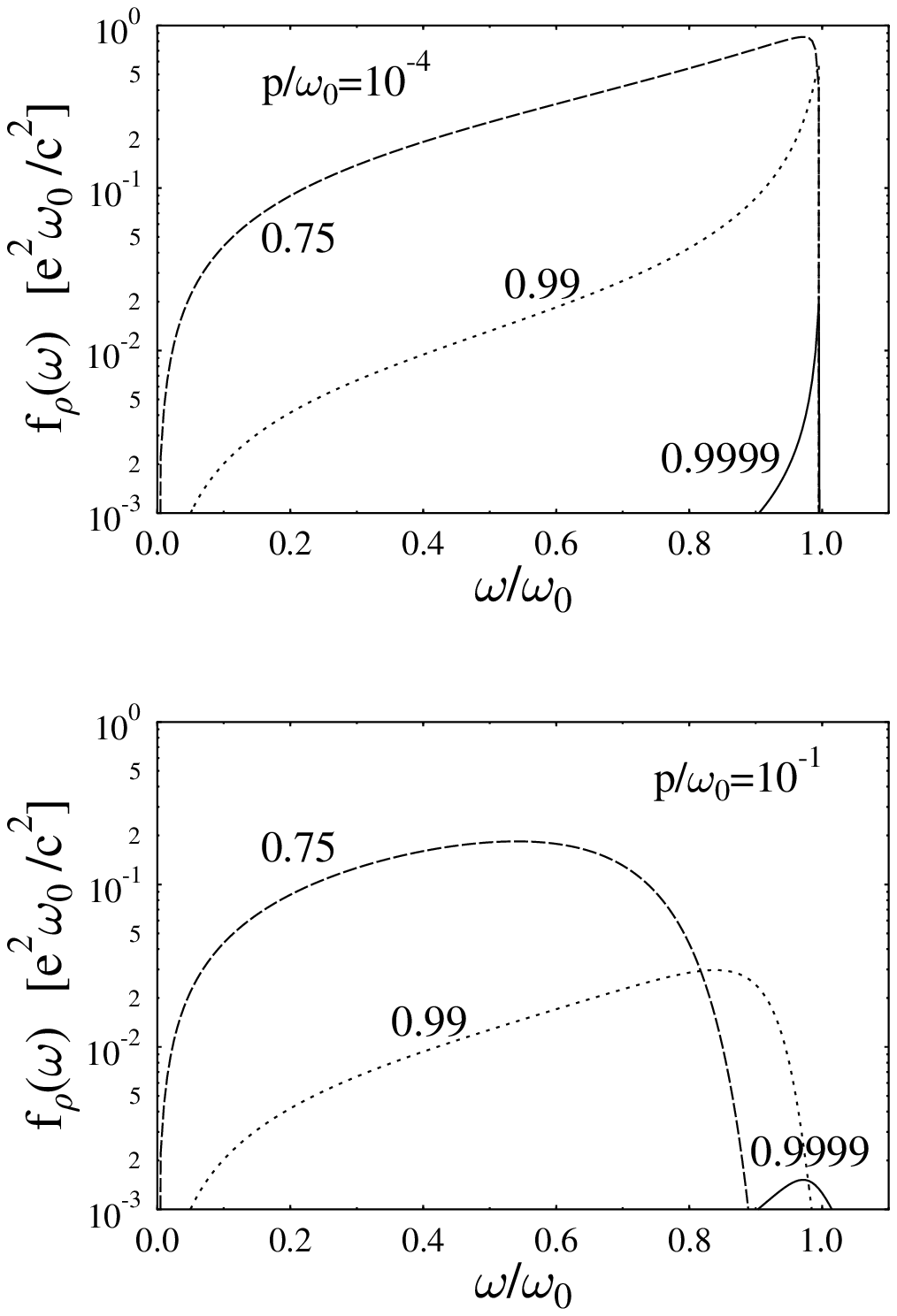}         
\end{center}
\vspace*{-5mm}
\caption{\small Spectral distributions of the radiated energy losses
per unit length of the cylinder C$_\rho$. The numbers on a particular
line means $\beta$ ($\beta_c$=0.999).}
\label{fig:fw}
\end{figure}

The spectral distributions $f_\rho(\omega)$ of the radiated energy
losses are shown in Fig.~\ref{fig:fw}.
The numbers of particular curves mean $\beta$.
It is seen  that for $\beta>\beta_c$ all $\omega$ from
the interval $0<\omega<\omega_0$ contribute to the energy losses.
For $\beta<\beta_c$ the interval of permissible $\omega$ diminishes.
The spectral density $f_\rho(\omega)$
of the radiated energy losses  defines the corresponding
photon number spectral density
$n_\rho(\omega) \equiv f_\rho(\omega)/\hbar\omega$
and the total number of photons emitted per unit length
$N_\rho = \int_{0}^{\infty} d\omega n_\rho(\omega)$.

In Fig.~\ref{fig:fw}, one can see that
the damping parameter reduces the
high frequency part $\omega\sim\omega_0$ of the domains
where VCR is emitted.
 Fig.~\ref{fig:fb} shows
that the damping strongly reduces the radiated energy losses
especially for the low velocities.
Spectral domains of VCR can approximately be fixed  by
the condition $Re(\epsilon)>\beta^{-2}$.
These domains are defined by the two
phenomenological parameters of the medium,
namely, the effective frequency $\omega_0$  of oscillators and
the damping
$p$ describing interactions with the other degrees of freedom.
The evaluated photon number spectral density $n_\rho(\omega)$
and the total number of photons emitted per unit length $N_\rho$
look very much like the corresponded spectral distributions
$f_\rho(\omega)$ and the radiated energy losses $W_\rho$,
respectively. To not overload the composition, we don't present
the figures for $n_\rho(\omega)$ and $N_\rho$.

It follows from Fig.~\ref{fig:sro} that
rapid oscillations of the radiation intensity as a function of time
should be observed in a particular detector.
To detect $ S_\rho$, one should have a detector imbedded into
a thin collimator and  directed towards the charge motion
axis. The collimator should be impenetrable for the $\gamma$ quanta
with directions different  from the  radial one.
The semiclassical WKB estimation~\cite{AKjpd98} for a period
$T\approx 2\pi/(\epsilon_0\omega_0)$ or the counting of the
peaks in Fig.~\ref{fig:sro} gives $T\sim~10^{-15}~sec$.
It seems that these oscillations shown in  Fig.~\ref{fig:sro}
for air-like media could be very
difficult to resolve experimentally.

\section{Conclusion}
We have evaluated electromagnetic potentials, strengths and
Poynting's vector describing space-time distribution of EMF
induced in the air-like medium by a fast charge particle
both below and above the Cherenkov threshold.

In the case of $\beta>\beta_c$
the response of a dispersive medium to a penetrating charge particle
manifests itself in the formation of a sharp front and rapid
wake-like oscillations behind the moving charge indicating
the appearance of electromagnetic shock waves.
A fine structure of EMF for $\beta<\beta_c$ manifests in
the existence of EMF radiation under
the Cherenkov threshold in the domain
of anomalous dispersion.

We have calculated the spectral distributions,
the number of the emitted photons and
the radiated EMF energy flux.
It is shown that the transition to the
subthreshold regime ($\beta>\beta_c\,\to\;\beta<\beta_c$)
and the damping ($p\neq 0$) lead to the sharp reduction of
the anomalous frequency range where VCR is emitted.

We acknowledge valuable discussions  with
Prof.~V.P.~Zrelov and Prof.~J.~Ruzi\~{c}ka.
Authors are thankful to Prof.~R.~Gupta for his constructive
criticisms of the manuscript.

\twocolumn


\end{document}